\documentclass[twocolumn,prl,groupedaddress,showpacs,amsfonts,amssymb,amsmath,letterpaper]{revtex4-1}
\usepackage{amsmath}
\usepackage{graphicx}
\usepackage{amssymb}

\makeatletter

\begin{document}

\title{
Persistent Quantum Beats and Long-Distance Entanglement from Waveguide-Mediated Interactions
}

\author{Huaixiu Zheng}
\email{hz33@duke.edu}
\author{Harold U. Baranger}
\email{baranger@phy.duke.edu}
\affiliation{\textit{Department of Physics, Duke University, P.O.\,Box 90305,
Durham, North Carolina 27708, USA}}

\date{Phys. Rev. Lett. \textbf{110}, 113601 (2013); March 12, 2013 }

\begin{abstract}
We study photon-photon correlations and entanglement generation in a
one-dimensional waveguide coupled to two qubits with an arbitrary spatial
separation. To treat the combination of nonlinear elements and 1D continuum, we
develop a novel Green function method. The vacuum-mediated qubit-qubit
interactions cause quantum beats to appear in the second-order
correlation function. We go beyond the Markovian regime and observe that such
quantum beats persist much longer than the qubit life time. 
A high degree of long-distance entanglement can be generated, 
increasing the potential of waveguide-QED systems for scalable quantum networking. 
\end{abstract}


\maketitle 

One-dimensional (1D) waveguide-QED systems are emerging
as promising candidates for quantum information processing
\cite{KimbleNat08,ChangPRL06,ChangNatPhy07,ShenPRL07,*ShenPRA07,ZhouPRL08,LongoPRL10,WitthautNJP10,ZhengPRA10,RephaeliPRA11,RoyPRL11,RoyPRA11,KolchinPRL11,ZhengPRL11,*ZhengPRA12,RephaeliPRL12},
motivated by tremendous experimental progress in a wide variety of
systems \cite{AkimovNat07, BajcsyPRL09, BabinecNatNanotech10,
ClaudonNatPhoton10, AstafievSci10,
AstafievPRL10,BleusePRL11,HoiPRL11,LauchtPRX12,HoiPRL12}. 
Over the past few years, a \emph{single} emitter strongly coupled to a 1D waveguide has been studied extensively
\cite{ChangPRL06,ChangNatPhy07,ShenPRA07,ShenPRL07,ZhouPRL08,LongoPRL10,WitthautNJP10,ZhengPRA10,RoyPRL11,KolchinPRL11,ZhengPRL11,ZhengPRA12,RephaeliPRL12}.
To enable greater quantum networking potential using waveguide-QED \cite{KimbleNat08}, it is important to study systems having more than just one qubit.

 \begin{figure}[b!]
 \centering
 \includegraphics[width=0.485\textwidth]{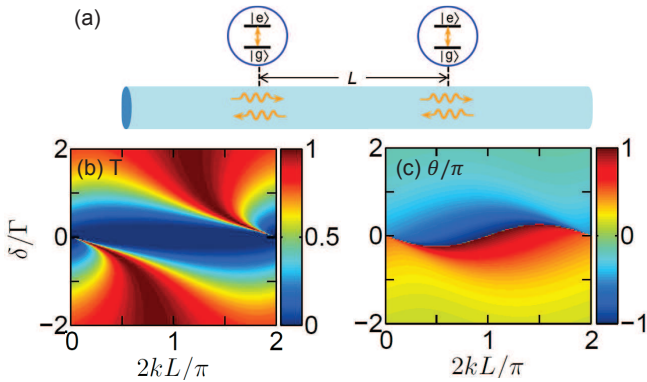}
\caption{Schematic diagram of the waveguide system and single-photon
transmission. (a) Two qubits (separated by $L$) interacting with the
waveguide continuum. Panels (b) and (c) show colormaps of the
single-photon transmission probability $T$ and the phase shift
$\theta$, respectively, as a function of detuning
$\delta=ck-\omega_0$ and $2kL$. 
Here, we consider the lossless case $\Gamma^{\prime}=0$.}
 \label{fig:Single_Photon}
 \end{figure}

In this Letter, we study cooperative effects of \emph{two} qubits strongly coupled to a 1D waveguide, finding the photon-photon correlations and qubit entanglement beyond the well-studied Markovian regime \cite{DzsotjanPRB10,DzsotjanPRB11,TudelaPRL11,CanoPRB11}. A key feature is the combination of these two highly nonlinear quantum elements with the 1D continuum of states. In comparison to either linear elements coupled to a waveguide \cite{PazPRL08,ZellPRL09,TanPRA11,WolfEPL11} or two qubits coupled to a single mode serving as a bus \cite{MajerNat07}, both of which have been studied previously, new physical effects appear. To study these effects, we develop a numerical Green function method to compute the photon correlation function for an arbitrary interqubit separation. 

The strong quantum interference in 1D, in contrast to the three-dimensional case \cite{FicekPhysRep02}, makes the vacuum-mediated qubit-qubit interaction \cite{DasPRL08} long-ranged. We find that quantum beats emerge in the photon-photon correlations, and persist to much longer time scales in the non-Markovian regime. We show that such persistent quantum beats arise from quantum interference between emission from two subradiant states. Furthermore, we demonstrate that a high-degree of long-distance entanglement can be generated, thus supporting
waveguide-QED-based open quantum networks.

\emph{Hamiltonian.\textemdash}As shown in Fig.\,1(a), we consider
two qubits with transition frequencies $\omega_1$ and $\omega_2$,
separation $L=\ell_2-\ell_1$, and dipole couplings to a 1D
waveguide. The Hamiltonian of the system is \footnote{Note that we
adopt the rotating wave approximation (RWA) at the level of
Hamiltonian. As pointed out in \cite{MilonniPRA95}, within the RWA
causality in photon propagation is preserved by extending the
frequency integrals to minus infinity. We carry out this scheme in
all of our numerical calculations.}
\begin{eqnarray}
H&=&\sum_{j=1,2}\hbar(\omega_j-i\Gamma_j^{\prime}/2)\sigma_j^{+}\sigma_j^- + H_{wg} \nonumber \\
&+&\sum_{j=1,2}\sum_{\alpha=R,L}\int dx\hbar V_j \delta(x-\ell_j)[a_{\alpha}^{\dagger}(x) \sigma_j^- +{\rm h.c.}] ,\nonumber \\
H_{wg}&=&\int dx\frac{\hbar c}{i}\Big[a_{R}^{\dagger}(x)\frac{d}{dx}a_{R}(x)-a_{L}^{\dagger}(x)\frac{d}{dx}a_{L}(x)\Big],
\label{eq:Ham}
\end{eqnarray}
where $a_{R,L}^{\dagger}(x)$ is the creation operator for a right-
or left-going photon at position $x$ and $c$ is the group velocity
of photons. $\sigma_j^{+}$ and $\sigma_j^- $ are the qubit raising
and lowering operators, respectively. An imaginary term in the
energy level is included to model the spontaneous emission of the
excited states at rate $\Gamma_{1,2}^{\prime}$ to modes other than
the waveguide continuum \cite{Carmichael93}. The decay rate to the
waveguide continuum is given by $\Gamma_j=2V_j^2/c$. Throughout the
Letter, we assume two identical qubits:
$\Gamma_1=\Gamma_2\equiv\Gamma$,
$\omega_1=\omega_2\equiv\omega_0\gg\Gamma$, and
$\Gamma_1^{\prime}=\Gamma_2^{\prime}\equiv\Gamma^{\prime}$.

\emph{Single-photon phase gate.\textemdash}
Assuming an incident photon from the left (with wave vector $k$), we obtain the single photon scattering eigenstate \footnote{see Supplementary Materials for details.}; the transmission coefficient is given by
\begin{eqnarray}
t_k\equiv\sqrt{T}e^{i\theta}=\frac{(ck-\omega_0+\frac{i\Gamma^{\prime}}{2})^2}{(ck-\omega_0+\frac{i\Gamma+i\Gamma^{\prime}}{2})^2+\frac{\Gamma^2}{4}e^{2ikL}}.
\label{eq:One-photon Transmission}
\end{eqnarray}
As shown in Fig.\,1(b), there is a large window of perfect
transmission: $T \!\approx\! 1$, even when the detuning
($\delta=ck-\omega_0$) of the single photon is within the resonance
line width ($\sim\!\Gamma$). This is in sharp contrast to the
single-qubit case, where perfect transmission is only possible for
far off-resonance photons \cite{ChangNatPhy07}. Such perfect
transmission occurs when the reflections from the two qubits
interfere destructively and cancel each other completely.
Furthermore, Fig.\,1(c) shows that within the resonance line width,
there is a considerable phase shift $\theta$. 
This feature of single-photon transmission can
be used to implement a photon-atom phase gate. For example, in the
case of $\delta=-0.5\Gamma$ and $kL=\pi/4$, the single photon passes
through the system with unit probability and a $\pi/2$ phase shift.
Two successive passes will give rise to a photon-atom $\pi$-phase
gate, which can be further used to realize a photon-photon phase
gate \cite{DuanPRL04}.

\emph{Photon-photon correlation: Nonlinear effects.\textemdash}To study 
the interaction effects, we develop a novel Green function method to calculate the 
full interacting scattering eigenstates and so photon-photon correlations.
We start with a reformulated Hamiltonian \cite{LongoPRL10}
\begin{eqnarray}
H&=&H_0+V,\,\qquad V=\sum_{j=1,2}\frac{U}{2}d_j^{\dagger}d_j(d_j^{\dagger}d_j-1), \nonumber \\
H_0&=&\sum_{j=1,2}\hbar(\omega_j-i\Gamma_j^{\prime}/2)d_j^{\dagger}d_j+H_{wg} \nonumber \\
   &+&\sum_{j=1,2}\sum_{\alpha=R,L}\int dx\hbar V_j \delta(x-a_j)[a_{\alpha}^{\dagger}(x) d_j+{\rm h.c.}],\quad
\label{eq:Ham_boson}
\end{eqnarray}
where $d_j^{\dagger}$ and $d_j$ are bosonic creation and
annihilation operators on the qubit sites. The qubit ground and
excited states correspond to zero- and one-boson states,
respectively. Unphysical multiple occupation is removed by including
a large repulsive on-site interaction term $U$; the Hamiltonians in
Eqs.\,(\ref{eq:Ham}) and (\ref{eq:Ham_boson}) become equivalent in
the limit $U\rightarrow\infty$. The non-interacting scattering
eigenstates can be obtained easily from
$H_0|\phi\rangle=E|\phi\rangle$. The full interacting scattering
eigenstates $|\psi\rangle$ are connected to $|\phi\rangle$ through
the Lippmann-Schwinger equation \cite{SakuraiQM94, DharPRL08,
RoyPRA11}
\begin{equation}
 |\psi\rangle = |\phi\rangle+G^{R}(E)V|\psi\rangle, \;\;\;
 G^{R}(E) = \frac{1}{E-H_0+i0^{+}}.
\label{Green_Fun}
\end{equation}
The key step is to numerically evaluate the Green functions, from which one obtains the scattering eigenstates \cite{Note2}.
Assuming a weak continuous wave incident laser, we calculate the second-order correlation function $g_2(t)$ \cite{LoudonQTL03} for an arbitrary interqubit separation.

Figure 2 shows $g_2(t)$ 
for both the transmitted and reflected fields 
when the probe laser is on resonance with the qubit: $k\!=\!k_0$ ($k_0\!\equiv\!\omega_0/c$).
When the two qubits are colocated \cite{RephaeliPRA11} ($L\!=\!0$), $g_2(t)$ of the transmitted field shows strong initial bunching followed by antibunching, while $g_2(t)$ of the reflected field shows perfect antibunching at $t\!=\!0$, $g_2(0)\!=\!0$.
This behavior is similar to that in the single qubit case \cite{ChangNatPhy07,ZhengPRA10}.
When the two qubits are spatially separated by $L=\pi/2k_0$, we observe 
quantum beats (oscillations). Since these beats occur in $g_2 (t)$, they necessarily involve the nonlinearity of the qubits and do not occur for, \textit{e.g.}, waveguide-coupled oscillators. 

As one increases the separation $L$, 
one may expect from the well-known 3D result that the quantum beats disappear \cite{FicekPRA90}. However, in our 1D system they do not: Figure 3 shows $g_2(t)$ for two cases, $k_0L=25.5\pi$ and $100.5\pi$, from which it is clear that the beats persist to long time. The 1D nature is key in producing strong quantum
interference effects and so long-range qubit-qubit interactions.

\begin{figure}[t]
 \centering
 \includegraphics[width=0.35\textwidth]{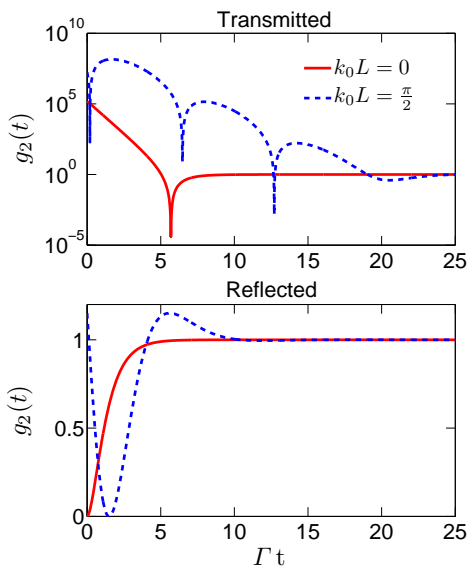}
 \caption{Quantum beats in the Markovian regime.
The se\-cond-order photon-photon correlation function of both the transmitted (top) and reflected (bottom) fields 
as a function of $t$ for $k_0L=0$ (solid line) and $k_0L=\pi/2$
(dashed line). The incident weak coherent state is on resonance with
the qubits: $k=k_0=\omega_0/c$. (Parameters: $\omega_0=100\Gamma$
and $\Gamma^{\prime}=0.1\Gamma$.)    }
 \label{fig:g2_Mark}
 \end{figure}

\emph{Non-Markovian regime.\textemdash}To interpret these exact numerical results, we compare them with the solution under the well-known Markov approximation. 
For small separations ($k_0L\leq\pi$), the system is Markovian \cite{FicekPRA90}: the causal propagation time of photons between the two qubits can be neglected and so the qubits interact instantaneously.
To understand quantum beats in this limit, we  use a master equation for the density matrix $\rho$ of the qubits in the Markov approximation. 
Integrating out the 1D bosonic degrees of freedom yields \cite{FicekPhysRep02}
\begin{eqnarray}
 \frac{\partial\rho}{\partial t}&=&\frac{i}{\hbar}[\rho,H_{c}]-\!\!\sum_{i,j=1,2}\frac{\Gamma_{ij}}{2}(\rho\sigma_i^{+}\sigma_j^- +\sigma_i^{+}\sigma_j^- \rho-2\sigma_i^- \rho\sigma_j^{+}),\nonumber \\
 H_{c}&=&\hbar\omega_0\sum_{i=1,2}\sigma_i^{+}\sigma_i^- +\hbar\Omega_{12}(\sigma_1^{+}\sigma_2^- +\sigma_2^{+}\sigma_1^- ),
\label{rho:Mark}
\end{eqnarray}
where $\Gamma_{ii}\equiv\Gamma+\Gamma^{\prime}$ while $\Gamma_{12}\equiv\Gamma \text{cos}(\omega_0L/c)$ and $\Omega_{12}\equiv(\Gamma/2) \text{sin}(\omega_0L/c)$ are the vacuum-mediated spontaneous and coherent couplings, respectively.
Transforming to symmetric and antisymmetric states $|S,A\rangle=(|g_1e_2\rangle\pm|e_1g_2\rangle)/\sqrt{2}$ gives a more transparent form:
\begin{eqnarray}
  \frac{\partial\rho}{\partial t}&=&\frac{i}{\hbar}[\rho,H_{c}]-\!\!\sum_{\beta=S,A}\frac{\Gamma_{\beta}}{2}(\rho\sigma_{\beta}^{+}\sigma_{\beta}^- +\sigma_{\beta}^{+}\sigma_{\beta}^- \rho-2\sigma_{\beta}^- \rho\sigma_{\beta}^{+}),\nonumber \\
 H_{c}&=&\sum_{\beta=S,A}\hbar\omega_{\beta}\sigma_{\beta}^{+}\sigma_{\beta}^- ,
\label{rhoSA:Mark}
\end{eqnarray}
where
$\sigma_{S,A}^{+}\equiv(\sigma_1^{+}\pm\sigma_2^{+})/\sqrt{2}$,
$\Gamma_{S,A}\equiv\Gamma+\Gamma^{\prime}\pm\Gamma_{12}$, and
$\omega_{S,A}\equiv\omega_0\pm\Omega_{12}$. 
Note that $|S\rangle$ and $|A\rangle$ are decoupled from each
other and have transition frequencies $\omega_{S,A}$ and decay rates
$\Gamma_{S,A}$ which oscillate as a function of $L$. When $L=0$,
$\Gamma_S=2\Gamma+\Gamma^{\prime}$ and $\Gamma_A=\Gamma^{\prime}$.
$|S\rangle$ is in the superradiant state, while $|A\rangle$ is
subradiant.
The waveguide couples only to the superradiant state and so the
photon-photon correlation mimics that for a
single-qubit. However, when $k_0L=\pi/2$,
$\Gamma_S=\Gamma_A=\Gamma+\Gamma^{\prime}$,
$\omega_{S,A}=\omega_0\pm\Gamma/2$, and the waveguide couples to
both $|S\rangle$ and $|A\rangle$. The quantum interference between
the transitions $|S\rangle\rightarrow|g_1g_2\rangle$ and
$|A\rangle\rightarrow|g_1g_2\rangle$ gives rise to quantum beats at 
frequency $\omega_S-\omega_A=\Gamma$, as shown in
Fig.\,\ref{fig:g2_Mark}.

\begin{figure}[t]
 \centering
 \includegraphics[width=0.35\textwidth]{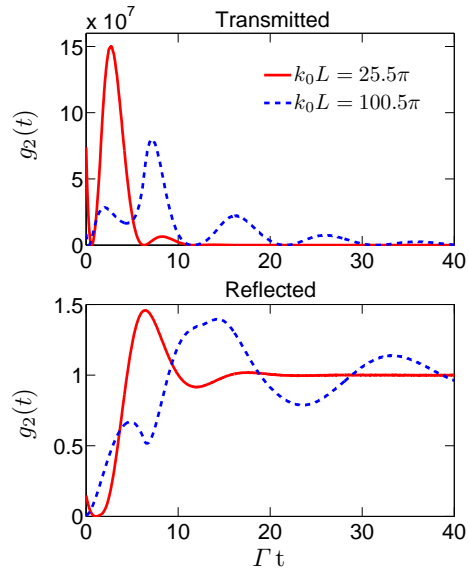}
 \caption{Persistent quantum beats in the non-Markovian regime.
The second-order correlation function of both the transmitted (top)
and reflected (bottom) fields is plotted as a function of $t$ for
$k_0L=25.5\pi$ (solid line) and $100.5\pi$ (dashed line). We set the
incident coherent state on resonance with the qubits ($k=k_0$),
$\omega_0=100\Gamma$ and $\Gamma^{\prime}=0.1\Gamma$.
\vspace*{-0.1in}}
 \label{fig:g2_NonMark}
 \end{figure}

As one increases the separation $L$ and goes beyond the Markovian
regime, Eq.\,(\ref{rho:Mark}) is not a valid description of the
system because the causal propagation time of photons (or retardation effect) has to be included. Comparing the results in Figs.\,\ref{fig:g2_Mark} and \ref{fig:g2_NonMark}, we see that quantum beats are \emph{more} visible in the non-Markovian regime in
both the transmitted and reflected fields and persist to a much
longer time scale, especially for the case $k_0L=100.5\pi$.

To better understand the persistent quantum beats, we extract the transition frequencies and decay rates of the two qubit system beyond the Markovian regime.
This is achieved by analyzing the poles of the Green function \cite{Note2} defined in Eq.\,(\ref{Green_Fun}); they are given by
\begin{equation}
 F(\omega)=\Big[\omega-\omega_0+\frac{i(\Gamma+\Gamma^{\prime})}{2}\Big]^2+\frac{\Gamma^2}{4}e^{2i\omega L/c}=0 \;.
\label{Poles}
\end{equation}
In the Markovian regime, one can safely replace $\omega$ by
$\omega_0$ in the exponent
, given that
$\omega_0\gg\Gamma$ and $L\ll c\Gamma^{-1}$. Eq.\,(\ref{Poles}) then yields
$\omega_{\pm}=\omega_0-i(\Gamma+\Gamma^{\prime})/2\pm
i\Gamma e^{i\omega_0L/c}/2$. The real and imaginary parts of
$\omega_{\pm}$ correspond to the transition frequencies and
decay rates, which are nothing but $\omega_{S,A}$ and
$-\Gamma_{S,A}/2$ obtained using the Markov approximation
[Eq.\,(\ref{rhoSA:Mark})]. Beyond this Markovian regime, we solve Eq.\,(\ref{Poles})
iteratively by gradually increasing $L$.

Figure 4 shows that both $\omega_{S,A}$ and
$\Gamma_{S,A}$ deviate significantly from their Markovian values as
$k_0L$ becomes large [Figs.\,4(c) and 4(d)].
The expanded detail plots, Figs.\,4(a) and
4(e), show that the Markov approximation works well for
$k_0L\in[0,5\pi]$. At large $k_0L$, however, \emph{both} the
symmetric and antisymmetric states become subradiant
[$\Gamma_{S,A}\ll\Gamma$, Fig.\,4(f)]. 
This suppression of decay comes about in the following way:
after the initial excitation of and emission from the first qubit, it can be reexcited by the pulse reflected from the second qubit. From the excitation probability of the first qubit through many emission-reexcitation cycles, an effective qubit life time can be defined: it is greatly lengthened by the causal propagation of photons between the two qubits.
$\Gamma_{S,A}$ characterize the average long time decay quantitatively.

 \begin{figure}[t!]
 \centering
 \includegraphics[width=0.48\textwidth]{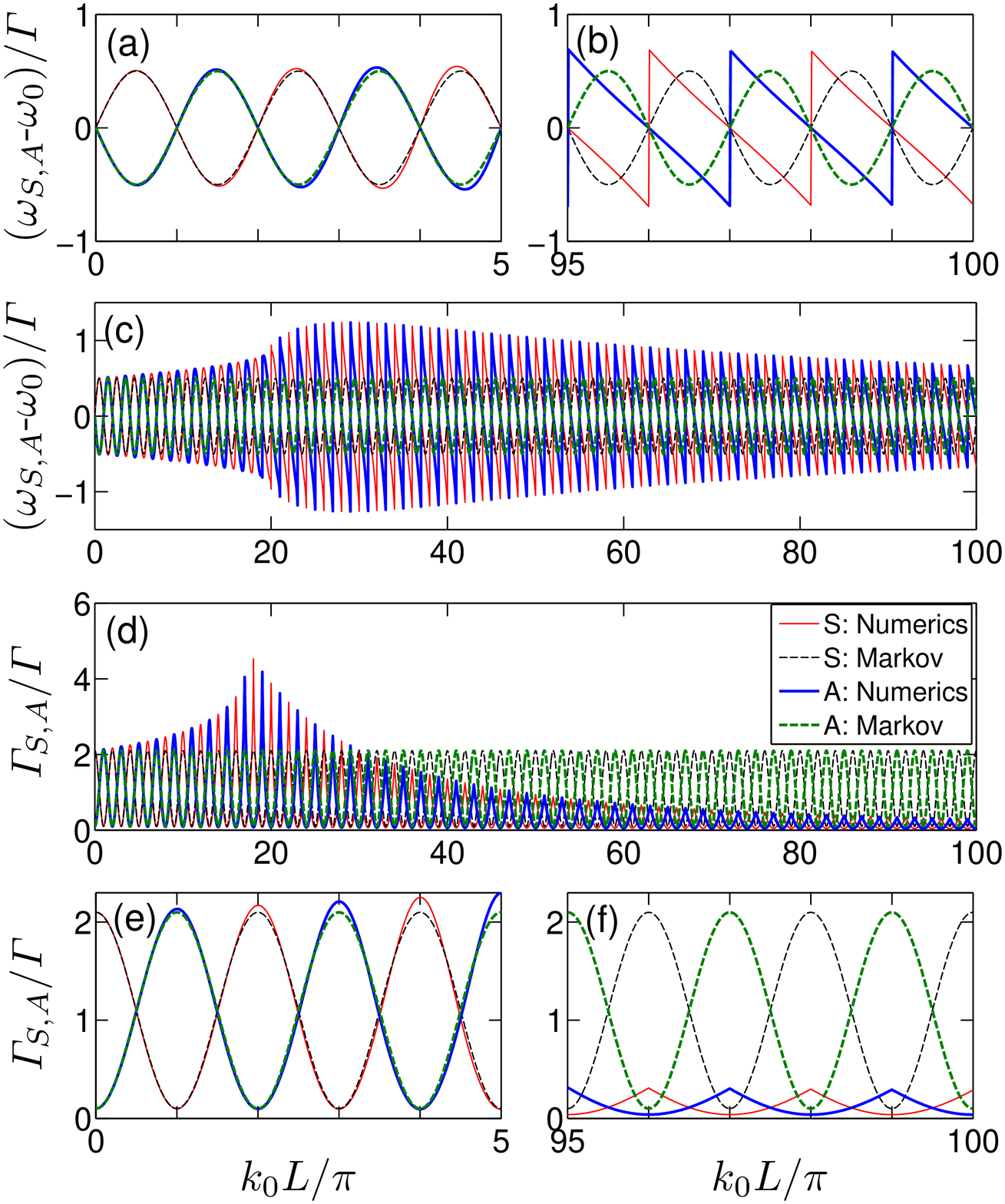}
 \caption{Renormalized transition frequencies and decay rates of the symmetric (S) and antisymmetric (A) states.
Panels (a)-(c) show the transition frequencies $\omega_{S}$ (thin
solid line) and $\omega_A$ (thick solid line) obtained numerically
from Eq.\,(\ref{Poles}) together with $\omega_{S}$ (thin dashed
line) and $\omega_A$ (thick dashed line) given by the Markov
approximation. Panels (d)-(f) similarly show the decay rates
$\Gamma_S$ and $\Gamma_A$ obtained both numerically and in the
Markov approximation. 
($\omega_0\!=\!100\Gamma$ and
$\Gamma^{\prime}\!=\!0.1\Gamma$.) }
 \label{fig:GAM_OM}
 \end{figure}

 \begin{figure}[t]
 \centering
 \includegraphics[width=0.35\textwidth]{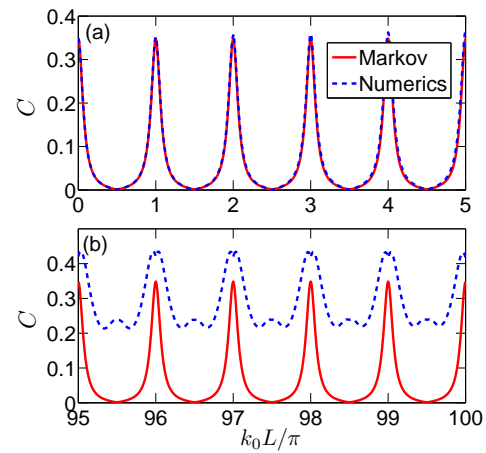}
 \caption{Long-distance qubit-qubit entanglement.
The steady state concurrence is plotted as a function of $k_0L$ for
(a) $0\leq k_0L\leq5\pi$, and (b) $95\pi\leq k_0L\leq100\pi$. The
Rabi frequencies are $\Omega_1=0.1\Gamma$, $\Omega_2=0$. The driving
laser is on resonance with the qubits. 
($\omega_0=100\Gamma$ and $\Gamma^{\prime}=0.1\Gamma$.) }
 \label{fig:Entanglement}
 \end{figure}

The nonlinear equation Eq.\,(\ref{Poles}) gives rise, of course, to infinitely many poles for $L>0$. These poles represent collective states of two spatially separated qubits with vacuum-mediated interactions. They are eigenmodes of the density matrix of the two qubits. The ``two-pole'' approximation of retaining only the symmetric and antisymmetric states is a good approximation because ($\omega_{S,A}-\omega_0,\Gamma_{S,A}$) are the two poles closest to the origin $(0,0)$. Within the parameter range we consider, all other collective states are far detuned from $\omega_0$ and hence barely populated \cite{Note2}. In addition, $|S\rangle$ and $|A\rangle$ have much smaller decay rates than all the other collective states. Therefore, these two slowly decaying states dominate the long-time dynamics and quantum interference between their spontaneous emissions is the physical origin of the persistent quantum beats observed in Fig.\,\ref{fig:g2_NonMark}. 

\emph{Qubit-qubit entanglement.\textemdash}With the two-pole
approximation, we study qubit-qubit entanglement using the
master equation Eq.\,(\ref{rhoSA:Mark}) with $\omega_{S,A}$ and
$\Gamma_{S,A}$ replaced by the renormalized values obtained from
Eq.\,(\ref{Poles}). We focus on the steady state case by including a
continuous weak driving laser on resonance with the first qubit:
$H_L=\hbar \Omega_1(\sigma_1^{+}+\sigma_1^- )$
\cite{TudelaPRL11,CanoPRB11}. The entanglement is characterized by
the concurrence \cite{WoottersPRL98}; Figure \ref{fig:Entanglement}
shows its steady state value for the Rabi frequency
$\Omega_1=0.1\Gamma$. For small separation
[Fig.\,5(a)], the concurrence agrees with that
obtained using the Markov approximation \cite{TudelaPRL11}: $C$
reaches its maximum 
when the maximally-entangled two-qubit subradiant state (either $|S\rangle$
or $|A\rangle$) has a minimal decay rate and is well populated
\cite{CanoPRB11}. Between two peaks, $C$ vanishes because
the symmetric and antisymmetric states are now barely populated and
the usual decay rate, $\Gamma+\Gamma^{\prime} \gg \Omega_1$, holds
\footnote{The population of an excited state with detuning $\Delta$,
decay rate $\Gamma$, and Rabi frequency $\Omega$ is given by
$1/[2+(\Delta/\Omega)^2+(\Gamma/2\Omega)^2]$.}.

In contrast, Fig.\,5(b) shows that the Markovian predictions break
down: we observe enhanced entanglement for an arbitrary interqubit
separation. Such enhancement is due to non-Markovian processes:  
both $|S\rangle$ and $|A\rangle$ become
subradiant (Fig.\,4) with decay rates much smaller than $\Gamma$ and hence are
well populated \cite{Note2}. Thus, 
long-range
entanglement is possible due to non-Markovian processes, making 1D
waveguide-QED systems promising candidates for scalable quantum
networking.

\emph{Discussion of loss.\textemdash}Accessing the 
non-Markovian regime requires a large (effective) distance between the qubits 
and hence low loss in the waveguide. Here, we have included the loss of the 
qubit by using an effective Purcell factor of 10 (\textit{i.e.}\ $\sim\!10\%$ loss). 
Because waveguide loss has the same effect on system performance as qubit loss (both lead to photon leakage), we expect that the observed persistent quantum beats and long-distance entanglement are robust against waveguide loss on this same level, namely $\sim\!10\%$. 
While some waveguides in current experimental systems are very lossy (such as plasmonic nanowires \cite{AkimovNat07}), we can circumvent this difficulty by using a hybrid nanofiber system as discussed in the Supplementary Material \cite{Note2}.
One example is an integrated fiber-plasmonic system \cite{ChangNatPhy07}: the optical fiber
is coupled to two tapered plasmonic nanowires which interact with local qubits (\textit{e.g.}\ quantum dots).
Another example is an integrated nanofiber-trapped atomic ensemble \cite{VetschPRL10, GobanPRL12}: 
an optical fiber is tapered into a nanofiber in two regions where atomic ensembles are trapped by the evanescent field surrounding the nanofibers.
In both of these examples, the long waveguide connecting the two qubits is a high quality optical fiber in which the loss is very small over a length of the order of $100$ wavelengths. 


We thank D.\ J.\ Gauthier for valuable discussions. This work was
supported by U.S. NSF\,Grant\,No.\,PHY-10-68698. H.Z.\ is
supported by a John T.\ Chambers Fellowship from the Fitzpatrick
Institute for Photonics at Duke Uni\-ver\-sity.
We thank the Fondation Nanosciences of Grenoble, France, for its hospitality during completion of this work. 

\bibliography{WQED_2012}

\newpage
\widetext

\begin{center}
\large\bf Supplementary Material for ``Persistent Quantum Beats and
Long-Distance Entanglement from Waveguide-Mediated Interactions''
\end{center}

\global\long\def\theequation{S\arabic{equation}}

\global\long\def\thefigure{S\arabic{figure}}
\setcounter{equation}{0}
\setcounter{figure}{0}

In this Supplementary Material we address the following topics: calculation of single-photon scattering eigenstates, our numerical Green function method, the two pole approximation, and possible low-loss systems for long-distance entanglement.

\bigskip

\noindent {\bf Single-Photon Scattering Eigenstates}\\
A general single-photon scattering eigenstate of the system described
by Eq.$\,$(1) in the main text reads

\begin{equation}
|\phi_{1}\rangle=\int dx\left[\phi_{R}(x)a_{R}^{\dagger}(x)+\phi_{L}(x)a_{L}^{\dagger}(x)+e_{1}\sigma_{1}^{+}+e_{2}\sigma_{2}^{+}\right]|0,g_{1}g_{2}\rangle,\label{eq:phi1}
\end{equation}

\noindent where $|0,g_{1}g_{2}\rangle$ is the zero photon state with
both qubits in the ground state. The Schr\'odinger equation $H|\phi_{1}\rangle=E|\phi_{1}\rangle$
gives

\begin{eqnarray}
\left[-i\hbar c\frac{d}{dx}-E\right]\phi_{R}(x)+\hbar V_{1}\delta(x-\ell_{1})e_{1}+\hbar V_{2}\delta(x-\ell_{2})e_{2} & = & 0,\nonumber \\
\left[i\hbar c\frac{d}{dx}-E\right]\phi_{L}(x)+\hbar V_{1}\delta(x-\ell_{1})e_{1}+\hbar V_{2}\delta(x-\ell_{2})e_{2} & = & 0,\nonumber \\
\left(\hbar\omega_{1}-i\Gamma_{1}^{\prime}/2-E\right)e_{1}+\hbar V_{1}\left[\phi_{R}(\ell_{1})+\phi_{L}(\ell_{1})\right] & = & 0,\nonumber \\
\left(\hbar\omega_{2}-i\Gamma_{2}^{\prime}/2-E\right)e_{2}+\hbar V_{2}\left[\phi_{R}(\ell_{2})+\phi_{L}(\ell_{2})\right] & = & 0.\label{eq:SchrodingerEq}
\end{eqnarray}

\noindent Assuming an incident right-going photon of wave vector $k=E/c$,
the wavefunction takes the following form

\begin{eqnarray}
\phi_{R}(x) & = & \frac{e^{ikx}}{\sqrt{2\pi}}\left[\theta(\ell_{1}-x)+t_{12}\theta(x-\ell_{1})\theta(\ell_{2}-x)+t_{k}\theta(x-\ell_{2})\right],\nonumber \\
\phi_{L}(x) & = & \frac{e^{-ikx}}{\sqrt{2\pi}}\left[r_{k}\theta(\ell_{1}-x)+r_{12}\theta(x-\ell_{1})\theta(\ell_{2}-x)\right],\label{eq:WF_Ansatz}
\end{eqnarray}

\noindent where $\theta(x)$ is the step function. Setting $\phi_{R,L}(\ell_{1,2})=[\phi_{R,L}(\ell_{1,2}^{+})+\phi_{R,L}(\ell_{1,2}^{-})]/2$
and plugging Eq.$\,$(\ref{eq:WF_Ansatz}) into (\ref{eq:SchrodingerEq}),
we obtain the following solution

\begin{eqnarray}
t_{12} & = & \frac{\left(ck-\omega_{1}+i\Gamma_{1}^{\prime}/2\right)\left(ck-\omega_{2}+i\Gamma_{2}^{\prime}/2+i\Gamma_{2}/2\right)}{\left(ck-\omega_{1}+i\Gamma_{1}^{\prime}/2+i\Gamma_{1}/2\right)\left(ck-\omega_{2}+i\Gamma_{2}^{\prime}/2+i\Gamma_{2}/2\right)+\Gamma_{1}\Gamma_{2}e^{2ikL}/4},\nonumber \\
r_{12} & = & \frac{-i\Gamma_{2}(ck-\omega_{1}+i\Gamma_{1}^{\prime}/2)e^{2ik\ell_{2}}/2}{\left(ck-\omega_{1}+i\Gamma_{1}^{\prime}/2+i\Gamma_{1}/2\right)\left(ck-\omega_{2}+i\Gamma_{2}^{\prime}/2+i\Gamma_{2}/2\right)+\Gamma_{1}\Gamma_{2}e^{2ikL}/4},\nonumber \\
t_{k} & = & \frac{\left(ck-\omega_{1}+i\Gamma_{1}^{\prime}/2\right)\left(ck-\omega_{2}+i\Gamma_{2}^{\prime}/2\right)}{\left(ck-\omega_{1}+i\Gamma_{1}^{\prime}/2+i\Gamma_{1}/2\right)\left(ck-\omega_{2}+i\Gamma_{2}^{\prime}/2+i\Gamma_{2}/2\right)+\Gamma_{1}\Gamma_{2}e^{2ikL}/4},\nonumber \\
r_{k} & = & \frac{-i\Gamma_{2}(ck-\omega_{1}+i\Gamma_{1}^{\prime}/2-i\Gamma_{1}/2)e^{2ik\ell_{2}}/2-i\Gamma_{1}(ck-\omega_{2}+i\Gamma_{2}^{\prime}/2+i\Gamma_{2}/2)e^{2ik\ell_{1}}/2}{\left(ck-\omega_{1}+i\Gamma_{1}^{\prime}/2+i\Gamma_{1}/2\right)\left(ck-\omega_{2}+i\Gamma_{2}^{\prime}/2+i\Gamma_{2}/2\right)+\Gamma_{1}\Gamma_{2}e^{2ikL}/4},\nonumber \\
e_{1} & = & \left(ic\sqrt{\frac{2}{\Gamma_{1}}}\right)\frac{e^{ik\ell_{1}}}{\sqrt{2\pi}}\left(t_{12}-1\right),\,\,\,\,\,\,\,\,\,\,\,\,\,\,\,\,\,\,\,\,\,\,\,\,\,\,\, e_{2}=\left(ic\sqrt{\frac{2}{\Gamma_{2}}}\right)\frac{e^{ik\ell_{2}}}{\sqrt{2\pi}}\left(t_{k}-t_{12}\right).\label{eq:tk_rk}
\end{eqnarray}

\noindent In the case of two identical qubits, $t_{k}$ reduces to
the expression given in Eq. (2) in the main text. 

Similarly, we can solve for the single-photon scattering eigenstate
for an incident right-going photon of wave vector $k=E/c$. We represent
the wavefunction with an incident right-going and left-going photon
by $|\phi_{1}(k)\rangle_{R}$ and $|\phi_{1}(k)\rangle_{L}$, respectively. \\

\noindent {\bf Numerical Green Function Method}\\
With the Lippmann-Schwinger equation shown in Eq.$\,$(4) of the main
text, we can solve for the full interacting solution. The non-interacting
eigenstates are simply products of single-photon states. 

\begin{eqnarray}
|\phi_{n}(k_{1},\cdots,k_{n})\rangle_{\alpha_{1},\cdots,\alpha_{n}} & = & |\phi_{1}(k_{1})\rangle_{\alpha_{1}}|\phi_{1}(k_{2})\rangle_{\alpha_{2}}\cdots|\phi_{1}(k_{n})\rangle_{\alpha_{n}},\,\,\alpha_{j}=R,\, L,\, j=1\text{-}n,\nonumber \\
H_{0}|\phi_{n}(k_{1},\cdots,k_{n})\rangle_{\alpha_{1},\cdots,\alpha_{n}} & = & c(k_{1}+\cdots+k_{n})|\phi_{n}(k_{1},\cdots,k_{n})\rangle_{\alpha_{1},\cdots,\alpha_{n}}.\label{eq:N_phi}
\end{eqnarray}

\noindent For simplicity, we will focus on the two-particle solution
from now on. Extending the formalism to the many-particle solution
is straightforward. The two-particle identity in real-space can be
written as 

\begin{eqnarray}
I_{2} & = & I_{2}^{x}\otimes|\emptyset\rangle\langle\emptyset|+I_{1}^{x}\otimes\sum_{i=1,2}|d_{i}\rangle\langle d_{i}|+I_{0}^{x}\otimes\sum_{i\leq j}|d_{i}d_{j}\rangle\langle d_{i}d_{j}|,\nonumber \\
I_{n}^{x} & = & \sum_{\alpha_{1}\cdots\alpha_{n}=R,L}\int dx_{1}\cdots dx_{n}|x_{1}\cdots x_{n}\rangle_{\alpha_{1}\cdots\alpha_{n}}\langle x_{1}\cdots x_{n}|,\label{eq:In}
\end{eqnarray}

\noindent where $|\emptyset\rangle$ is the ground state of the two
qubits (bosonic sites), $|d_{i}\rangle=d_{i}^{\dagger}|\emptyset\rangle$,
$|d_{i}d_{i}\rangle=\frac{(d_{i}^{\dagger})^{2}}{\sqrt{2}}|\emptyset\rangle$
and $|d_{1}d_{2}\rangle=d_{1}^{\dagger}d_{2}^{\dagger}|\emptyset\rangle$.
Inserting the above identity into Eq. (4) in the main text, we obtain

\begin{eqnarray}
|\psi_{2}(k_{1},k_{2})\rangle_{\alpha_{1},\alpha_{2}} & = & |\phi_{2}(k_{1},k_{2})\rangle_{\alpha_{1},\alpha_{2}}+G^{R}(E)VI_{2}|\psi_{2}(k_{1},k_{2})\rangle_{\alpha_{1},\alpha_{2}}\nonumber \\
 & = & |\phi_{2}(k_{1},k_{2})\rangle_{\alpha_{1},\alpha_{2}}+UG^{R}(E)\sum_{i=1,2}|d_{i}d_{i}\rangle\langle d_{i}d_{i}|\psi_{2}(k_{1},k_{2})\rangle_{\alpha_{1},\alpha_{2}}.\label{eq:phi2}
\end{eqnarray}

\noindent Projecting Eq.$\,$(\ref{eq:phi2}) onto $\langle d_{i}d_{i}|$
yields

\begin{equation}
\left(\begin{array}{c}
\langle d_{1}d_{1}|\psi_{2}(k_{1},k_{2})\rangle_{\alpha_{1},\alpha_{2}}\\
\langle d_{2}d_{2}|\psi_{2}(k_{1},k_{2})\rangle_{\alpha_{1},\alpha_{2}}
\end{array}\right)=\left(\begin{array}{c}
\langle d_{1}d_{1}|\phi_{2}(k_{1},k_{2})\rangle_{\alpha_{1},\alpha_{2}}\\
\langle d_{2}d_{2}|\phi_{2}(k_{1},k_{2})\rangle_{\alpha_{1},\alpha_{2}}
\end{array}\right)+U\left[\begin{array}{cc}
G_{11} & G_{12}\\
G_{21} & G_{22}
\end{array}\right]\left(\begin{array}{c}
\langle d_{1}d_{1}|\psi_{2}(k_{1},k_{2})\rangle_{\alpha_{1},\alpha_{2}}\\
\langle d_{2}d_{2}|\psi_{2}(k_{1},k_{2})\rangle_{\alpha_{1},\alpha_{2}}
\end{array}\right),\label{eq:Phi2_d1d1}
\end{equation}

\noindent where we introduce the short-hand notation $G_{ij}=\langle d_{i}d_{i}|G^{R}(E)|d_{j}d_{j}\rangle$.
Solving Eq. (\ref{eq:Phi2_d1d1}) gives rise to

\begin{equation}
\left(\begin{array}{c}
\langle d_{1}d_{1}|\psi_{2}(k_{1},k_{2})\rangle_{\alpha_{1},\alpha_{2}}\\
\langle d_{2}d_{2}|\psi_{2}(k_{1},k_{2})\rangle_{\alpha_{1},\alpha_{2}}
\end{array}\right)=\left(I-U\left[\begin{array}{cc}
G_{11} & G_{12}\\
G_{21} & G_{22}
\end{array}\right]\right)^{-1}\left(\begin{array}{c}
\langle d_{1}d_{1}|\phi_{2}(k_{1},k_{2})\rangle_{\alpha_{1},\alpha_{2}}\\
\langle d_{2}d_{2}|\phi_{2}(k_{1},k_{2})\rangle_{\alpha_{1},\alpha_{2}}
\end{array}\right).\label{eq:Phi2_G}
\end{equation}

\noindent Projecting Eq.$\,$(\ref{eq:phi2}) onto a two-photon basis
state $\langle x_{1}x_{2}|$ and taking the $U\rightarrow\infty$
limit, we obtain the full interacting two-photon solution

\begin{eqnarray}
\langle x_{1}x_{2}|\psi_{2}(k_{1},k_{2})\rangle_{\alpha_{1},\alpha_{2}} & = & \langle x_{1}x_{2}|\phi_{2}(k_{1},k_{2})\rangle_{\alpha_{1},\alpha_{2}}+U\left(\begin{array}{cc}
G_{1}(x_{1},x_{2}) & G_{2}(x_{1},x_{2})\end{array}\right)\left(\begin{array}{c}
\langle d_{1}d_{1}|\psi_{2}(k_{1},k_{2})\rangle_{\alpha_{1},\alpha_{2}}\\
\langle d_{2}d_{2}|\psi_{2}(k_{1},k_{2})\rangle_{\alpha_{1},\alpha_{2}}
\end{array}\right)\nonumber \\
 & = & \langle x_{1}x_{2}|\phi_{2}(k_{1},k_{2})\rangle_{\alpha_{1},\alpha_{2}}-G_{xd}G_{dd}^{-1}\left(\begin{array}{c}
\langle d_{1}d_{1}|\phi_{2}(k_{1},k_{2})\rangle_{\alpha_{1},\alpha_{2}}\\
\langle d_{2}d_{2}|\phi_{2}(k_{1},k_{2})\rangle_{\alpha_{1},\alpha_{2}}
\end{array}\right),\label{eq:Two_photon_solution}
\end{eqnarray}

\noindent where $G_{i}(x_{1},x_{2})=\langle x_{1}x_{2}|G^{R}(E)|d_{i}d_{i}\rangle$
and 

\begin{eqnarray}
G_{xd} & \equiv & \left(\begin{array}{cc}
G_{1}(x_{1},x_{2}) & G_{2}(x_{1},x_{2})\end{array}\right),\nonumber \\
G_{dd} & \equiv & \left[\begin{array}{cc}
G_{11} & G_{12}\\
G_{21} & G_{22}
\end{array}\right].\label{eq:G_matrix}
\end{eqnarray}

Hence, the remaining task is to calculate all the Green functions
in Eq. (\ref{eq:Two_photon_solution}). This can be done using the
two-photon non-interacting scattering eigenstates, from which we can
construct a two-particle identity in momentum space 

\begin{equation}
I_{2}^{\prime}=\sum_{\alpha_{1},\alpha_{2}=R,L}\int dk_{1}dk_{2}|\phi_{2}(k_{1},k_{2})\rangle_{\alpha_{1},\alpha_{2}}\langle\phi_{2}(k_{1},k_{2})|.\label{eq:I2_q}
\end{equation}

\noindent Using Eq. (\ref{eq:N_phi}), the Green functions can be
evaluated as

\begin{eqnarray}
G_{ij} & = & \langle d_{i}d_{i}|G^{R}(E)I_{2}^{\prime}|d_{j}d_{j}\rangle\nonumber \\
 & = & \sum_{\alpha_{1},\alpha_{2}=R,L}\int dk_{1}dk_{2}\frac{1}{E-ck_{1}-ck_{2}+i0^{+}}\langle d_{i}d_{i}|\phi_{2}(k_{1},k_{2})\rangle_{\alpha_{1},\alpha_{2}}\langle\phi_{2}(k_{1},k_{2})|d_{j}d_{j}\rangle,\nonumber \\
G_{i}(x_{1},x_{2}) & = & \langle x_{1}x_{2}|G^{R}(E)I_{2}^{\prime}|d_{i}d_{i}\rangle\nonumber \\
 & = & \sum_{\alpha_{1},\alpha_{2}=R,L}\int dk_{1}dk_{2}\frac{1}{E-ck_{1}-ck_{2}+i0^{+}}\langle x_{1}x_{2}|\phi_{2}(k_{1},k_{2})\rangle_{\alpha_{1},\alpha_{2}}\langle\phi_{2}(k_{1},k_{2})|d_{i}d_{i}\rangle.\label{eq:Gij_Gi}
\end{eqnarray}

\noindent Doing the integrals numerically gives the full interacting
two-particle solution. Again, following the same program, it is straightforward
to extend the formalism to the three- or more photon solution with
two or more qubits coupled to the waveguide. \\

\begin{figure}[tb!]
 \centering
 \includegraphics[width=0.98\textwidth]{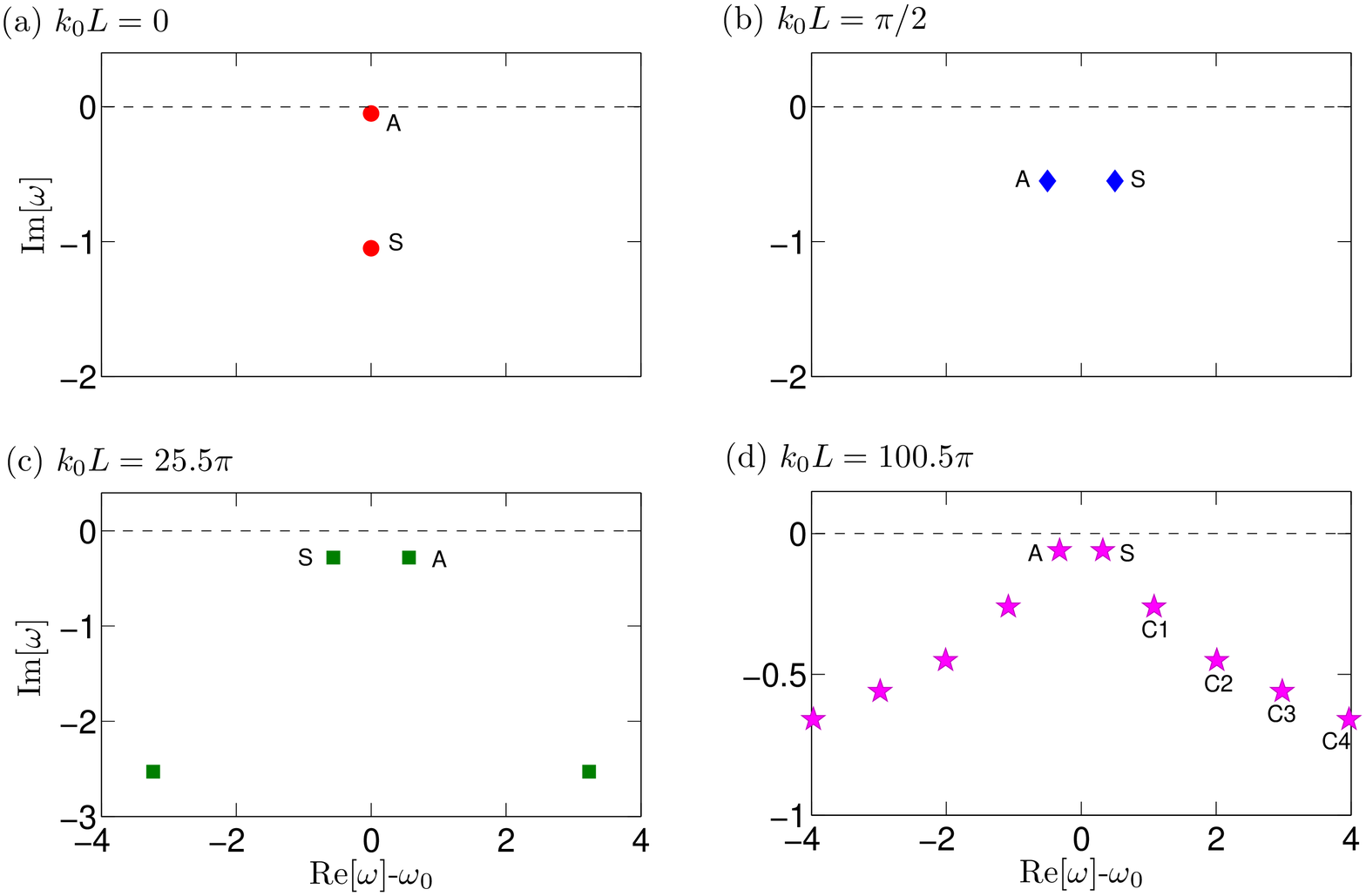}
\caption{The poles of the Green functions for (a) $k_{0}L=0$, (b) $k_{0}L=\frac{\pi}{2}$,
(c) $k_{0}L=25.5\pi$, and (d) $k_{0}L=100.5\pi$. Both the real and
imaginary parts of the poles are in units of $\Gamma$. We show all
the poles with real part within $[\omega_0-4\Gamma,\,\omega_0+4\Gamma]$. The poles
corresponding to the $|S\rangle$ and $|A\rangle$ states are labeled
as $S$ and $A$, respectively. In case (d), there are four additional
poles ($C1\text{-}C4$) within the plotted range.}
\end{figure}

\noindent {\bf Two-Pole Approximation}\\
In this section, we will show the validity of the `two-pole' approximation
in the parameter regime we consider. Assuming two identical qubits,
the poles of the Green functions in Eq. (\ref{eq:Two_photon_solution}) are given
by

\begin{equation}
F(\omega)=\left[\omega-\omega_{0}+\frac{i(\Gamma+\Gamma^{\prime})}{2}\right]^{2}+\frac{\Gamma^{2}}{4}e^{2i\omega L/c}=0.\label{eq:Poles}
\end{equation}

\noindent Figure S1 plots the poles computed numerically in four different
cases. For small $L$, Figs.$\,$S1(a) and S1(b) show that there are
only two poles corresponding to $|S\rangle$and $|A\rangle$ states
within a large range of frequency. 
At large $L$, however, both the
symmetric and antisymmetric states become subradiant
[$\Gamma_{S,A}\ll\Gamma$]. 
This suppression of decay comes about in the following way:
after the initial excitation of and emission from the first qubit, it can be reexcited by the pulse reflected from the second qubit. From the excitation probability of the first qubit through many emission-reexcitation cycles, an effective qubit life time can be defined: it is greatly lengthened by the causal propagation of photons between the two qubits.
$\Gamma_{S,A}$ characterize the average long time decay quantitatively.
Furthermore, as $L$ increases, there
are additional poles as shown in Figs.$\,$S1(c) and S1(d), corresponding
to collective states generated in non-Markovian processes. 
For $L\gg c\Gamma^{-1}$, the two-pole approximation breaks down as the additional poles of collective states become close enough to $|S\rangle$and $|A\rangle$ states.

Here, we want to analyze the case $k_{0}L=100.5\pi$ [Fig.\,S1(d)], where $L\sim c\Gamma^{-1}$ and the two-pole approximation is still valid as we will show below. With
a driving laser on resonance with the qubits and a Rabi frequency
$\Omega$, the probability to excite a state $|y\rangle$ $(\omega_{y},\,\Gamma_{y})$
is

\begin{equation}
P_{y}=\frac{1}{2+\left(\frac{\omega_{y}-\omega_{0}}{\Omega}\right)^{2}+\left(\frac{\Gamma_{y}}{2\Omega}\right)^{2}}.\label{eq:Py}
\end{equation}

\noindent Using this formula, we can calculate the probability of
exciting the states corresponding to $S\,(\omega_{0}+0.32\Gamma,0.12\Gamma)$,
$A\,(\omega_{0}-0.32\Gamma,0.12\Gamma)$, $C1\,(\omega_{0}+1.08\Gamma,0.52\Gamma)$
, $C2\,(\omega_{0}+2.01\Gamma,0.90\Gamma)$, $C3\,(\omega_{0}+2.98\Gamma,1.12\Gamma)$
and $C4\,(\omega_{0}+3.97\Gamma,1.32\Gamma)$. In the limit of weak
driving laser, $\Omega\rightarrow0$, we have

\begin{eqnarray}
P_{C1} & = & 8.6\%P_{S},\nonumber \\
P_{C2} & = & 2.5\%P_{S},\nonumber \\
P_{C3} & = & 1.1\%P_{S},\nonumber \\
P_{C4} & = & 0.7\%P_{S}.\label{eq:PC1-C4}
\end{eqnarray}

\noindent Hence, compared to states $C1\text{-}C4$, $S$ and
$A$ states are well populated and dominate the qubit-qubit interactions
for the parameter regime considered in the main text. \\

\noindent {\bf Possible Low-Loss Systems for Long-Distance Entanglement}\\
In this section, we discuss the issue of waveguide loss and propose several low-loss systems to overcome this difficulty.
As discussed in the main text, waveguide loss has to be limited to the same level as qubit loss. 
However, some waveguides in current experiments, \textit{e.g.}\ plasmonic nanowires, are too lossy to meet this criteria. 
We propose to use either a hybrid optical fiber systems or slow-light superconducting systems.
In the first case, low-loss optical fibers are used to transmit light over a long distance. 
The transmission length we are considering is of order 100 wavelengths, thus of order 100 microns for typical quantum dots or atoms. 
Loss over such a distance in state of the art fiber is very small: taking a 4dB/km fiber, the loss will be on the order of 1\,ppm. 
In the second case, the actual transmission length is very short, but due to the reduced speed of light one can still reach the non-Markovian regime. Below are three plausible experimental settings: (a) and (b) belong to the first case and (c) illustrates the second 
case.
\medskip

\begin{figure}[tb!]
 \centering
 \includegraphics[width=0.7\textwidth]{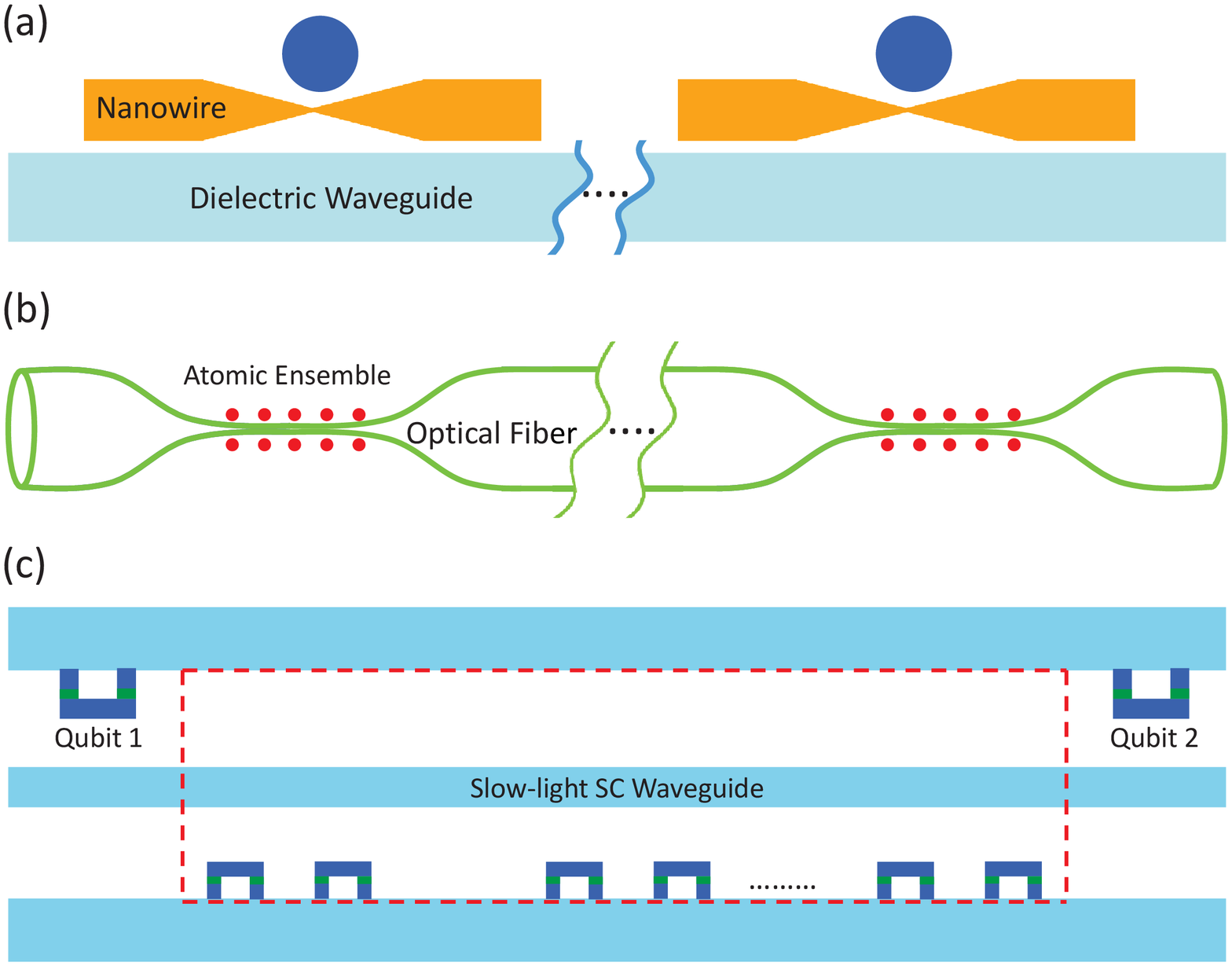}
\caption{Three possible setups for long-distance entanglement in waveguide-QED. 
(a) Hybrid plasmonic system. The dielectric waveguide (light blue) is phase-matched with the plasmonic nanowires (yellow) so that efficient plasmon transfer between them can be realized. 
The nanowires are strongly coupled to the quantum dots (blue) in the tapered regions. Note that the length of the dielectric waveguide between the nanowires can be very long (indicated by breaks).
(b) Tapered nanofiber system. An optical fiber (green) is tapered into narrow nanofibers in two regions. Two atomic ensembles (red) are trapped by and strongly coupled to the nanofibers.
(c) Slow-light superconducting system. Two superconducting qubits (blue) couple strongly to the slow-light superconducting waveguide (red dashed box).
}
\label{fig:setups}
\end{figure}

\noindent \emph{(a) Hybrid Fiber-Plasmonic Waveguide-QED System} \\
Figure \,\ref{fig:setups}(a) shows an integrated fiber-plasmonic waveguide-QED system. 
The idea of hybrid plasmonic systems was first proposed by Chang
et al. \cite{ChangNatPhy07}. Since then, there has been extensive experimental \cite{BensonNat11,Tiecke12}, and theoretical \cite{OultonNatPhoton08,Feist12,ZouIEEE12} work along this
line. In the schematic, the optical fiber is coupled to two tapered plasmonic nanowires. Due to the subwavelength confinement \cite{ChangNatPhy07}, the
plasmonic field in the nanowires couples strongly to the local qubits, \textit{e.g.}\ quantum dots \cite{AkimovNat07}. 
Coupling the nanowires to a dielectric waveguide ensures that the quantum state can be transmitted over long distance without being dissipated in the nanowires.

\medskip

\noindent \emph{(b) Integrated Nanofiber-Trapped Atomic Ensemble System} \\
In the second example, a long optical fiber is tapered into a narrow nanofiber in two regions. Then, two atomic ensembles are trapped by the evanescent field surrounding the nanofibers.
Strong coupling is achieved between the propagating photons in the nanofiber and the atomic ensembles \cite{VetschPRL10}.
Such a setting is a clear extension of the experimental systems demonstrated by several groups \cite{VetschPRL10, GobanPRL12}.

\medskip

\noindent \emph{(c) Slow-light Superconducting Waveguide-QED System} \\
In the third example, a 1D open superconducting transmission line is coupled to two superconducting qubits. 
It has been experimentally demonstrated that this system is deep in the strong coupling regime \cite{HoiPRL11}.
However, the typical length of the transmission line is on the order of the wavelength of propagating microwave photons. 
Hence, the separation of the two qubits is limited to the photon wavelength. 
To reach the non-Markovian regime, we can make the effective distance between the two qubits large by the slow-light scheme first proposed by Shen and Fan \cite{ShenPRB07b}.
The idea is to couple the transmission line to an additional periodic array of unit cells made of two qubits. 
Flat photonic bands can be generated to slow down the microwave photons.
While not true long-distance propagation, this could be an effective way to experimentally probe non-Markovian effects.

\end{document}